\documentclass{aipproc}
\layoutstyle{8x11double}

\begin{document}

\title{Phase Structure of Confining Theories on $R^3$ x $S^1$}


\author{Hiromichi Nishimura}{
  address={Department of Physics, Washington University, St.\ Louis, Missouri 63130, USA}
}

\author{Michael C. Ogilvie}{
  address={Department of Physics, Washington University, St.\ Louis, Missouri 63130, USA}
}

\begin{abstract}

Recent work on QCD-like theories on $R^3\times S^1$ has revealed that a confined phase can exist when the circumference $L$ of $S^1$ is sufficiently small. Adjoint QCD and double-trace deformation theories with certain conditions are such theories, and we present some new results for their phase diagrams. First we show the connection between the large-$L$ and small-$L$ confined regions in the phase diagram of $SU(3)$ adjoint QCD using Polyakov-Nambu-Jona Lasinio models. Then we consider an $SU(2)$ double-trace deformation theory with adjoint scalars and study conflicts between the Higgs and small-$L$ confined phase. 

\end{abstract}

\maketitle

Recent progress in QCD-like theories has shown that confinement can be understood analytically using semi-classical methods when one or more spatial directions are compactified and small \cite{Unsal:2007vu,Myers:2007vc,Unsal:2008ch,Unsal:2007jx}.
Because the perturbative contribution of the gauge fields to the effective potential always favors the deconfined phase at a small circumference, \emph{e.g.}, the deconfinement transition in finite-temperature gauge theories, it is necessary to modify the gauge theory in some way to obtain confinement with small compact directions. There are currently two ways to realize the small-$L$ confining phase: a double-trace deformation theory which directly modifies the gauge action with terms nonlocal in the compact directions and adjoint QCD with periodic boundary conditions in the compact directions. 

In this paper, we extend and clarify the phase structures of adjoint QCD and double-trace deformation theories on $R^3\times S^1$.
First, we consider the connection between the small-$L$ and large-$L$ confined regions in the phase diagrams.
Interestingly, there are intermediate phases between the confined and deconfined phases at a small compactified circumference for $SU(N)$ with $N\ge3$  \cite{Myers:2007vc,Myers:2009df}.
In double-trace deformation theories, it is known from lattice simulations that the small-$L$ confining phase of $SU(3)$ is continuously connected to the conventional large-$L$ confining phase \cite{Myers:2007vc}.
For the case of adjoint QCD, the finite mass of adjoint fermions suppresses the fermionic contribution to the effective potential, so the effect of chiral symmetry breaking may be important in obtaining small-$L$ confinement. In order to explore the interrelationship of confinement and chiral symmetry breaking, we use a generalization of Nambu-Jona Lasinio (NJL) models known as Polyakov-Nambu-Jona Lasinio (PNJL) models \cite{Fukushima:2003fw} and extend it to the case of adjoint fermions \cite{Nishimura:2009me}.
Secondly, we introduce adjoint scalar fields in a double-trace deformation theory. The aim is to set up a potential conflict between confinement and the Higgs mechanism. We briefly mention the role of topological excitations in each phase in this theory. 

Our principle tool to determine the phase diagram is the effective potential in terms of the order parameters of the specific model under consideration. 
The perturbative calculation of the effective potential is reliable at a small circumference $L$ for the compact direction because QCD-like theories are asymptotically free. For the scalar sector, we assume that the running coupling for the quartic scalar interaction is small at the relevant mass scale of the theory. 
The order parameter for confinement-deconfinement is the Polyakov loop
\begin{equation}
P\left(\vec{x}\right)=\mathcal{P}\exp\left[i\int_{0}^{L}dx_{4}A_{4}\left(x\right)\right]
\end{equation}
which transforms nontrivially under center symmetry $Z(N)$. In adjoint QCD, $\bar{\psi} \psi$ is the order parameter for chiral symmetry breaking. The scalar field $\phi$ is the order parameter for symmetry breaking in the scalar sector, but it is gauge-variant. 

\section{Adjoint QCD and PNJL Models}

We extend PNJL models to the case of $SU(3)$ gauge theory with $N_f=2$ of Dirac fermions in the adjoint representation \cite{Nishimura:2009me}.
For gauge bosons, the one-loop free energy in a
background Polyakov loop is given by
\begin{equation}
V_{GL}=2\, Tr_{A}[\frac{1}{L}\int\frac{d^{3}k}{(2\pi)^{3}}ln(1-Pe^{-L\Omega_{k}})]
\end{equation}
where we have inserted a mass parameter in $\Omega_{k}=\sqrt{k^{2}+M^{2}}$
for purely phenomenological reasons \cite{Meisinger:2001cq}. 
A small-$L$ expansion gives the correct one-loop energy independent of the mass parameter $M$, and by setting $M=596\,MeV$, the next-order term proportional to $M^2/L^2$ yields the correct deconfinement temperature for the pure gauge theory, with a value of $T_d\approx270\,MeV$.

The fermionic part of the Lagrangian of our PNJL model is similar to that of NJL model. The $L^{-1}=0$ contribution to the effective potential, $V_{F0}\left(g_S,\Lambda,m\right)$ is unchanged. The potential depends on a four-fermion coupling constant $g_S$, a noncovariant three-dimensional cutoff $\Lambda$, and a constituent mass $m=-2g_S\langle \bar{\psi}\psi\rangle$ with zero current mass. In the PNJL model, the $L$-dependent part of the one-loop fermionic contribution to the effective potential is
\begin{equation}
V_{FL}=-2N_fTr_{A}[\frac{1}{L}\int\frac{d^{3}k}{(2\pi)^{3}}ln(1- Pe^{-L\omega_k})+h.c.]
\end{equation}
where $\omega_k=\sqrt{k^2+m^2}$. This term offsets the gluonic potential, maintaining $Z(N)$ symmetry provided that $mL$ is sufficiently small. 

\begin{figure}
  \includegraphics[width=.356\textheight]{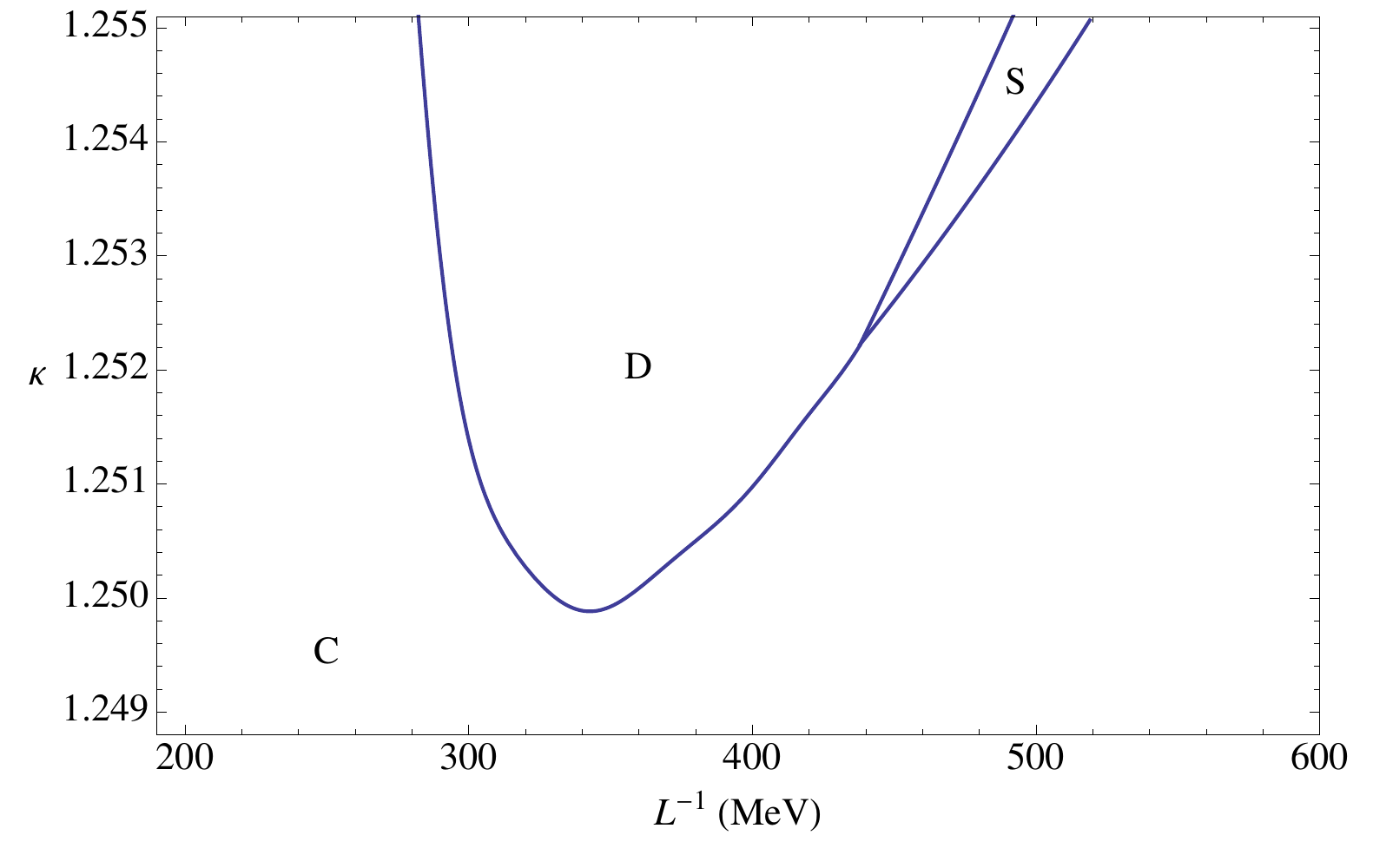}
  \caption{The phase diagram for two-flavor adjoint
QCD with periodic boundary conditions
in the $L^{-1}$-$\kappa$ plane.
C, D and S refer to the confined phase, deconfined phase,
and skewed phase, respectively.}
\end{figure}

For the case of fundamental fermions, our model reproduces the known crossover phase transitions similar to the original PNJL of Fukushima \cite{Fukushima:2003fw}. For the case of adjoint fermions, we calibrate our model with antiperiodic boundary conditions to fix the parameters in $V_{F0}$. 
We take the ratio $m\left(L^{-1}=0\right)/\Lambda$ to be $0.1$, which is small enough for the cutoff theory to make sense, and set $\Lambda=23.22\,GeV$ so that the ratio of chiral symmetry restoration temperature $T_c$ to the deconfinement temperature $T_d$ becomes $T_c/T_d \approx 7.8$, which is known from the lattice simulations \cite{Karsch:1998qj,Engels:2005te}.
We use the same value of $\Lambda$ for the case of periodic boundary conditions, assuming that boundary conditions do not change the scale of cutoff in the theory. 

We minimize the total effective potential with respect to two order parameters, the Polyakov loop $P$ and the constituent mass $m$ as a function of the dimensionless coupling constant, $g_S\Lambda^2\equiv\kappa$. As a result, we obtain a phase diagram in the $L^{-1}-\kappa$ plane for the $SU(3)$ gauge theory with two adjoint Dirac fermions as shown in Figure 1. This phase diagram is compatible with the lattice simulation by Cossu and D'Elia \cite{Cossu:2009sq}. However, the PNJL model shows that the small-$L$ and large-$L$ confined regions are connected for lower values of $\kappa$, which correspond to lower values of constituent mass for $L^{-1}=0$. Only for a very small value of $\kappa$, chiral symmetry seems to be restored with the condition $L^{-1}<<\Lambda$, but it is difficult to resolve in our calculations.

\section{Double-Trace Deformation and Higgs Mechanism}

We now consider an SU(2) adjoint Higgs theory on $R^{3}\times S^{1}$. The conventional part of the Euclidean action is given by
\begin{equation}
S_{c}=\int d^{4}x\left[\frac{1}{4}\left(F_{\mu\nu}^{a}\right)^{2}+\frac{1}{2}\left(D_{\mu}\phi\right)^{T}\cdot D_{\mu}\phi+V\left(\phi\right)\right]
\end{equation}
where $V\left(\phi\right)=1/2m^2\phi^2+1/4\lambda\left(\phi^2\right)^2$ and $D_{\mu}\phi=\partial_{\mu}\phi-igA_{\mu}\phi$. 
In addition, we need a double-trace deformation term $V_{dt}$ in order to realize the confined phase for small $L$. Many forms of $V_{dt}$ may be used, but here we choose the form
\begin{equation}
V_{dt}=\frac{a}{2\pi^{2}L^{4}}\sum_{n=1}^{\infty}\frac{\left|Tr_{F}P^{n}\right|^{2}}{n^{2}}.
\end{equation}
The infinite series can be summed exactly, leading to an analytically tractable expression for the effective potential. This deformation leads to a second-order deconfinement transition at some critical value $a_c$. 
The action has two global  symmetries: a $Z(2)_H$ symmetry in which $\phi \rightarrow -\phi$ and a $Z(2)_C$ center symmetry in which $P \rightarrow -P$. There are three distinct gauge-invariant order parameters associated with the  symmetries; $\left\langle Tr_{F}P\left(x\right)\right\rangle $, which transforms non-trivially under $Z(2)_C$; $\left\langle Tr_{F}\left[P^2\left(x\right)\phi(x)\right]\right\rangle $, which transforms non-trivially under $Z(2)_H$; and $\left\langle Tr_{F}\left[P\left(x\right)\phi(x)\right]\right\rangle $, which transforms non-trivially under both groups.

\begin{figure}
  \includegraphics[height=.21\textheight]{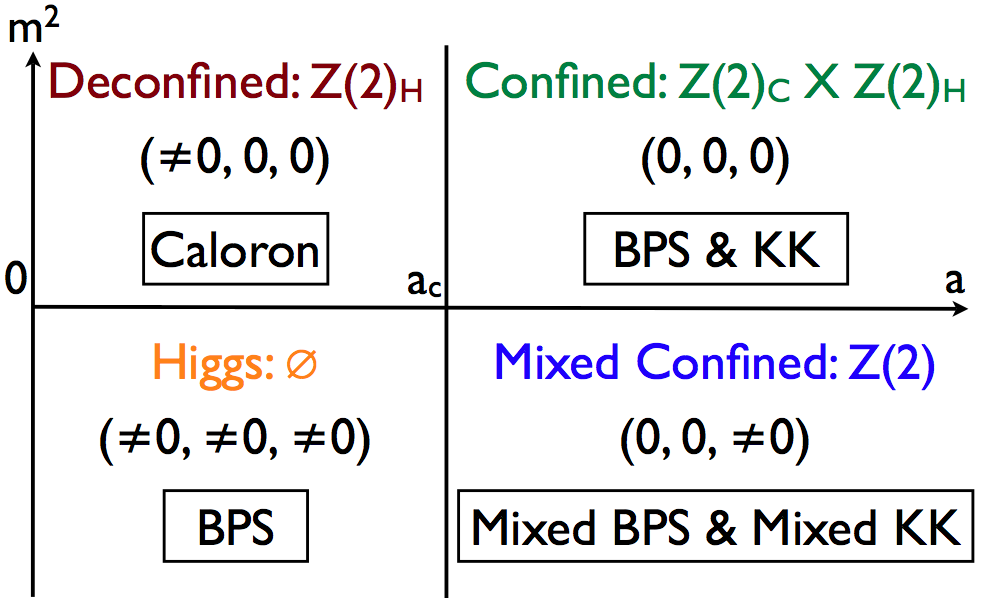}
  \caption{The phase diagram for the double-trace deformation theory with adjoint scalars in $a-m^2$ plane with the topological objects identified in each phase. The numbers are the values of gauge-invariant order parameters, $\left( \left\langle Tr_{F}P\left(x\right)\right\rangle ,\left\langle Tr_{F}\left[P^{2}\left(x\right)\phi(x)\right]\right\rangle ,\left\langle Tr_{F}\left[P\left(x\right)\phi(x)\right]\right\rangle \right)$.}
\end{figure}

We use the following simple approximation to describe the overall phase structure: we will include all tree-level contributions to the effective potential, including the term $V_{dt}$, and the leading $L^{-1}$ term from one loop. Thus our approximate form for the effective potential is simply
\begin{equation}
V_{eff}\simeq Tr\left[A_{4},\phi\right]^{2}+V\left(\phi\right)+\left[V_{GL}+V_{\phi L}+V_{dt}\right]\left(P\right)
\end{equation}
where $V_{\phi L}$ is the one-loop potential for $\phi$, which is half of $V_{GL}$ because it is spinless.
Note that the first term in $V_{eff}$ does nothing but force $A_4$ and $\phi$ to lie in the same direction in the $SU(2)$ Lie algebra. We choose a gauge where the Polyakov loop lies in the $3$-direction of $SU(2)$, independent of $\vec{x}$, so we can write the value of $P$ as $P=\exp\left(i\theta\sigma_3\right)$ and $\phi$ as $\phi=v\sigma_{3}$, where $\sigma_i$ is the Pauli matrix. Minimization of the effective potential reduces to minimizing two separate functions of a single variable: $V\left(\phi\right)$ and $V_{GL}\left(P\right)+V_{\phi L}\left(P\right)+V_{dt}\left(P\right)$. 
The expected values $\theta$ and $v$ are not themselves gauge-invariant, but they can be used reliably to calculate the gauge-invariant order parameters mentioned above for small $L$. 
In this simplified approximation, $v$ is zero for $m^{2}>0$ and non-zero for $m^{2}<0$. On the other hand, for $a>a_c$, $\theta$ is $\pi/2$ and the $Z(2)_C$-symmetric, confined phase is favored. For $a<a_c$, $\theta \neq \pi/2$, and $Z(2)_C$ is spontaneously broken.    
Neglected terms in the full one-loop potential couple $\phi$ and $P$ directly, and shift the phase diagram somewhat from the predictions of this simple approximation. However, they do not change the overall nature of the phase diagram.

We summarize our results in a phase diagram shown in Figure 2. There are four possible spontaneous symmetry breaking patterns of $Z(2)_C\times Z(2)_H$. Thus, with the confining phase corresponding to unbroken $Z(2)_C\times Z(2)_H$, there seem to be five possible phases. However, the order parameters do not allow a phase where both the Higgs mechanism and confinement hold, characterized by $\left( \left\langle Tr_{F}P\left(x\right)\right\rangle ,\left\langle Tr_{F}\left[P^{2}\left(x\right)\phi(x)\right]\right\rangle ,\left\langle Tr_{F}\left[P\left(x\right)\phi(x)\right]\right\rangle \right)=(0,\neq0,0)$ with only a $Z(2)_C$ symmetry unbroken. This is consistent with the general results of 't Hooft \cite{'tHooft:1979uj}.
On the other hand, there is a phase where $Z(2)_C\times Z(2)_H$ spontaneously breaks to $Z(2)$. 
We call this phase a mixed confined phase because symmetry breaking is realized by a linear combination of $\phi$ and $A_4$. There are topological objects of mixed BPS monopoles and mixed Kaluza-Klein (KK) monopoles, whose actions are $4\pi/g^{2}\sqrt{4\theta^{2}+g^{2}L^{2}v^{2}}$ and $4\pi/g^{2}\sqrt{4\left(\pi-\theta\right)^{2}+g^{2}L^{2}v^{2}}$, respectively. If $Z(2)_H$ is restored, then $v=0$, and they reduce to the actions of usual BPS and KK, which are the constituents of calorons \cite{Lee:1997vp,Kraan:1998pm}. 

\section{Conclusions}

We have presented some new results for the phase diagrams in the QCD-like theories on $R^3\times S^1$.
For $SU(3)$ adjoint QCD with two Dirac fermions with periodic boundary conditions, we have extended a Polyakov-Nambu-Jona Lasinio model, which incorporates both chiral symmetry breaking and confinement, to the case of adjoint fermions. The phase diagram in this model is compatible with the lattice simulation by Cossu and D'Elia. The large-$L$ and small-$L$ confined regions are connected in the phase diagram for a sufficiently small constituent mass.

For $SU(2)$ double-trace deformation theories with adjoint scalar fields, we have shown that according to the gauge-invariant order parameters, there is no phase where small-$L$ confinement and the Higgs mechanism take place. We have found a new mixed confined phase, where $Z(2)_C\times Z(2)_H \rightarrow Z(2)$ is realized by two Higgs fields $\phi$ and $A_4$. We have also constructed monopole solutions in the mixed confined phase using a linear combination of $\phi$ and $A_4$.

\begin{theacknowledgments}
  The authors thank the U.S. Department of Energy for financial support. 
\end{theacknowledgments}

\bibliographystyle{aipproc}   
\bibliography{MadridConfinement}

\end{document}